\documentclass[aps,prd,twocolumn,groupedaddress]{revtex4}

\usepackage{graphicx}% Include figure files

\usepackage{dcolumn}% Align table columns on decimal point

\usepackage{bm}% bold math

\renewcommand\){\right)}

\newcommand{\ra}{\rightarrow}

\def\lsim{\raise 0.4ex\hbox{$<$}\kern -0.8em\lower 0.62
ex\hbox{$\sim$}}

\def\gsim{\raise 0.4ex\hbox{$>$}\kern -0.7em\lower 0.62
ex\hbox{$\sim$}}

\def\lbar{{\hbox{$\lambda$}\kern -0.7em\raise 0.6ex
\hbox{$-$}}}

\newcommand\eq[1]{eq.~(\ref{#1})}
\newcommand\eqs[2]{eqs.~(\ref{#1}) and (\ref{#2})}

\newcommand\p{\partial}

\newcommand\ee{\end{equation}}
\newcommand\be{\begin{equation}}
\def\bea{\begin{array}}
\def\eea{\end{array}}\def\ea{\end{array}}
\newcommand\ees{\end{eqnarray}}
\newcommand\bees{\begin{eqnarray}}

\def\p1{{\bf p}_1}
\def\p2{{\bf p}_2}
\def\k1{{\bf k}_1}
\def\k2{{\bf k}_2}

\newcommand{\dddM}{\kern 0.2em \raise 1.9ex\hbox{$...$}\kern -1.0em \hbox{$M$}}
\newcommand{\dddQ}{\kern 0.2em \raise 1.9ex\hbox{$...$}\kern -1.0em \hbox{$Q$}}
\newcommand{\dddI}{\kern 0.2em \raise 1.9ex\hbox{$...$}\kern -1.0em\hbox{$I$}}
\newcommand{\dddJ}{\kern 0.2em \raise 1.9ex\hbox{$...$}\kern-1.0em
\hbox{$J$}}
\newcommand{\dddcalJ}{\kern 0.2em \raise 1.9ex\hbox{$...$}\kern-1.0em
\hbox{${\cal J}$}}

\newcommand{\dddO}{\kern 0.2em \raise 1.9ex\hbox{$...$}\kern -1.0em
\hbox{${\cal O}$}}
\def\dddz{\raise 1.5ex\hbox{$...$}\kern -0.8em \hbox{$z$}}
\def\dddd{\raise 1.8ex\hbox{$...$}\kern -0.8em \hbox{$d$}}
\def\dddbd{\raise 1.8ex\hbox{$...$}\kern -0.8em \hbox{${\bf d}$}}
\def\ddbd{\raise 1.8ex\hbox{$..$}\kern -0.8em \hbox{${\bf d}$}}
\def\dddx{\raise 1.6ex\hbox{$...$}\kern -0.8em \hbox{$x$}}

\newcommand{\Sch}{Schwarzschild }

\def\D{\Delta}
\def\p{\partial}
\def\a{\alpha}

\def\g{\gamma}

\def\l{\lambda}

\def\d{\delta}

\def\dslash{\hspace{-1mm}\not{\hbox{\kern-2pt $\partial$}}}
\def\Dslash{\not{\hbox{\kern-4pt $D$}}}
\def\pslash{\not{\hbox{\kern-2.1pt $p$}}}
\def\kslash{\not{\hbox{\kern-2.3pt $k$}}}
\def\qslash{\not{\hbox{\kern-2.3pt $q$}}}

\newcommand{\lpl}{l_{\rm Pl}}

\newcommand{\inT}{\int_{-\infty}^{\infty}}

\begin{document}

%\preprint{UGVA-DPT 2003/xxx}

\title{The physical interpretation of the spectrum \\
of black hole quasinormal modes}

\author{Michele Maggiore}
\affiliation{D\'epartement de Physique Th\'eorique, 
Universit\'e de Gen\`eve, 24 quai Ansermet, CH-1211 Gen\`eve 4}

%\date{\today}

\begin{abstract}
When a classical black hole is perturbed, its relaxation is governed by a set
of quasinormal modes with complex frequencies $\omega=
\omega_R+i\omega_I$. We show that this behavior is the same as 
that of  damped harmonic
oscillators whose
real frequencies are $(\omega_R^2+\omega_I^2)^{1/2}$,
 rather than simply $\omega_R$.
Since, for highly excited modes,
$\omega_I\gg \omega_R$, this observation
changes drastically the physical understanding of
the black hole spectrum, and forces a reexamination of various results in the
literature. In particular, adapting a derivation 
by Hod, we find that the area of
the horizon of a \Sch black hole
is quantized in units $\D A=8\pi\lpl^2$, in contrast with the
original result  $\D A=4\log(3) \lpl^2$.

\end{abstract}



%\keywords{}


\maketitle

Perturbations of black holes (BHs) vanish in time as a superposition of damped
oscillations, of the form 
\be\label{oRoT}
 e^{-\omega_I t}[a \sin (\omega_R t) + b \cos (\omega_R t)]\, ,
\ee
with a spectrum of
complex frequencies $\omega =\omega_R+i\omega_I$.  
These quasinormal modes are of great importance in
gravitational-wave astrophysics, and might be observed in existing or advanced
gravitational-wave detectors. Furthermore, BHs are often used as a testing
ground for ideas in quantum gravity, and their quasinormal modes are obvious
candidates for an interpretation in terms of quantum levels.

\begin{figure}
\includegraphics[width=0.4\textwidth]{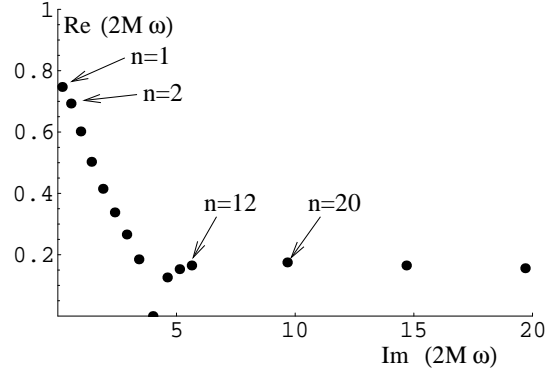}
\caption{\label{fig1} ${\rm Re} (2M \omega_{nl})$ against ${\rm Im}
(2M\omega_{nl})$ for $l=2$ and
$n=1,2,\ldots 12$, and for $n=20,30,40$. Data taken from \cite{Berti}. }
\end{figure}

For \Sch BHs, the quasinormal mode frequencies
are labeled as $\omega_{nl}$, where $l$ is the angular momentum quantum number.
For each $l$ (with $l\geq 2$ for gravitational perturbation), 
there is
a countable infinity of quasinormal 
modes, labeled by the ``overtone'' number $n$, with
$n=1,2,\ldots $.
In Fig.~\ref{fig1} we show the frequencies of the $l=2$ 
gravitational perturbations of a  \Sch BH of mass $M$:
${\rm Im}\,\omega_n$ grows monotonically with
$n$, so the least damped mode corresponds to $n=1$, and has
$2M {\rm Im}\,\omega \simeq 0.1779$ (we use units $G=c=1$). This
is the mode that dominates the
relaxation process. The next least damped mode is $n=2$, with 
$2M {\rm Im}\,\omega \simeq 0.5478$, and so on. In contrast, 
the real part of $\omega$
is not monotonic with $n$. It rather decreases at first, until it becomes 
exactly zero
for $n=9$, and then starts 
growing again, reaching a constant asymptotic value.
For large $n$ the asymptotic behavior of the  frequencies
of gravitational perturbations is independent of $l$, and is given 
by~\cite{Nollert:1999ji,Leaver,Noll93,Hod,Motl,Motl:2003cd,AndH,Visser,pad1}
\be\label{asy}
8\pi M\omega_{n}=\ln 3 + 2\pi i \(n+\frac{1}{2}\) +O\(n^{-1/2}\)\, .
\ee

The  pattern shown in Fig.~\ref{fig1}
repeats for higher $l$. There is always a value $\bar{n}_l$
of $n$ such that, for $n<\bar{n}_l$, ${\rm Re} (\omega_{n,l})$ 
decreases with $n$, while above this critical value
it raises again, up to the  asymptotic value  $\ln 3/(8\pi M)$
given by \eq{asy}.

If we compare  with the normal mode structure  of familiar
classical systems, such as a vibrating rod, we have to
admit that the  structure displayed in Fig.~\ref{fig1}, and particularly the
``inverted branch'' formed by the modes with
$n\leq \bar{n}_l$, is quite
peculiar. In classical systems, the least damped mode is in general
also the one with the
{\em lowest} value of ${\rm Re}\,\omega$,  and typically 
${\rm Re}\,\omega$ and  ${\rm Im}\,\omega$  both increase with $n$.
In contrast, in  Fig.~\ref{fig1}
 the least damped mode is 
the one with the
{\em highest} possible value of ${\rm Re}\,\omega$ and, 
for $n<\bar{n}_l$, 
${\rm Re}\,\omega$ is a decreasing function of 
${\rm Im}\,\omega$!
Even the ``normal'' branch $n>\bar{n}_l$ is somewhat puzzling. Now 
${\rm Re}\,\omega$ increases with $n$, which is more consistent with physical
intuition, but still the fact that it saturates to a finite value is difficult
to understand. In a normal macroscopic system, the underlying reason why, for
large $n$, ${\rm Im}\,\omega_n$ goes to infinity
(and therefore these modes decay very fast),
is that also ${\rm Re}\,\omega_n$ diverges, so increasing $n$ the wavelength 
$\l_n =2\pi/{\rm Re}\,\omega_n$ gets smaller and smaller, 
and finally becomes of the same order as the
lattice spacing of the underlying atomic structure. At this point the
perturbation can no longer be sustained as a wave by the medium, and quickly 
disappears
in the thermal agitation of the lattice nuclei.

The quasinormal mode structure of Fig.~\ref{fig1} is no less puzzling if we
attempt a semiclassical description and we interprete
it as the structure of excited levels of a quantum BH.
In normal quantum systems, the
levels with high excitation energy, $E_n=\hbar {\rm Re}\,\omega_n$, 
are those that 
decay fast, first of all because, in a multipole expansion, the decay width
$\Gamma$ grows  with $\omega$ (e.g. $\Gamma\sim \omega^3$ 
for a dipole transition
and $\Gamma\sim \omega^5$ for a quadrupole transition) and, second, 
because they can decay into  many different
channels, i.e. into all the levels with lower
excitation energy not forbidden by selection rules. 
So, again it is very surprising  that, for $n<\bar{n}_l$, we have an
inverted structure, where the lifetime of the state {\em increases} with its
excitation energy.  Quite puzzling is also the presence of a state with 
${\rm Re}\,\omega_n=0$, and ${\rm Im}\,\omega_n\neq 0$ 
(which exists for all $l$). 
So, the motivation of this work was to try to obtain a physical
understanding of this level structure.

To this purpose, we describe
a quasinormal mode  as
a damped harmonic oscillator $\xi (t)$, governed by the equation
\be
\ddot{\xi} +\g_0\dot{\xi}+\omega_0^2\xi =f(t)\, ,
\ee
where $\g_0$ is the damping constant, $\omega_0$ the proper frequency of the
harmonic oscillator, and $f(t)$ an
external force per unit mass. Solving this equation in Fourier transform we get
\be\label{xiopm}
\xi(t) =-\inT\, \frac{d\omega}{2\pi}\, 
\frac{\tilde{f}(\omega)}{(\omega-\omega_+)(\omega-\omega_-)}\, e^{i\omega t}
\, ,
\ee
where $\omega_{\pm}$ are the two roots of the equation 
$\omega^2-i\g_0\omega -\omega_0^2=0$,
i.e.
\be
\omega_{\pm}=\pm \sqrt{\omega_0^2-(\g_0/2)^2}\, +i\frac{\g_0}{2}\, .
\ee
Consider  the response to a Dirac delta perturbation, $f(t)\propto
\d (t)$, so $\tilde{f}(\omega)\propto 1$. 
For $t<0$ we can close the integration
contour in \eq{xiopm} in the lower half-plane and, since $\omega_{\pm}$ both
lie above the real axis, we get zero, as required by causality. For $t>0$ we
close the contour in the upper half-plane and we pick the residue of the two
poles. So the result for $\xi (t)$ is a superposition of a term
oscillating as $e^{i\omega_+t}$ and of a term oscillating as
$e^{i\omega_-t}$. Therefore, the behavior (\ref{oRoT}) is 
reproduced by a damped
harmonic oscillator, with the identifications
\be
\frac{\g_0}{2}=\omega_I\, ,\hspace{5mm} 
\sqrt{\omega_0^2-(\g_0/2)^2}=\omega_R\, ,
\ee
which can be inverted to give
\be\label{o0oRoI}
\omega_0=\sqrt{\omega_R^2+\omega_I^2}\, .
\ee
We see that the seemingly obvious  
identification $\omega_0=\omega_R$ only holds  
when $\g_0/2\ll\omega_0$, i.e. for
very long-lived modes. For most BH quasinormal modes we are in 
the opposite limit; 
in particular, for highly
excited modes, we  have $\omega_I\gg\omega_R$, see Fig.~\ref{fig1}, so
$\omega_0\simeq \omega_I$ rather than $\omega_0\simeq \omega_R$!
If we  model the  BH perturbations in terms of a collection of damped
harmonic degrees of freedom (which can be useful both classically, to have an
intuitive physical picture of a BH as a whole, and in semiclassical
quantum gravity, to get hints about the quantum structure of spacetime)
the correct identification for the frequency of the equivalent harmonic
oscillator is given by \eq{o0oRoI}, together with $\g_0/2=\omega_I$.

\begin{figure}
\includegraphics[width=0.4\textwidth]{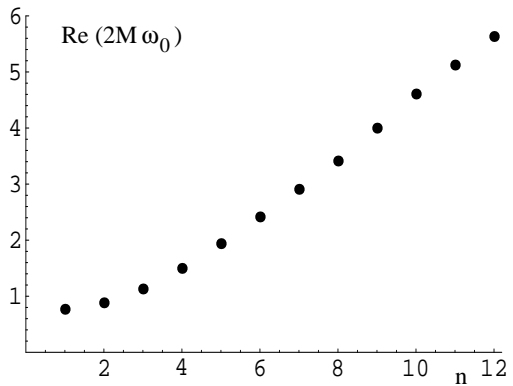}
\caption{\label{fig2} ${\rm Re} [2M (\omega_0)_{nl}]$ 
for $l=2$, against $n$.}
\end{figure}

In terms of $\omega_0$
the energy level structure of a BH 
becomes
 physically very reasonable, and for $l=2$
it is shown in Fig.~\ref{fig2}
(a similar result holds for higher $l$). We see that
the frequency $(\omega_0)_n $ increases monotonically with the overtone
number $n$. Recall that the damping coefficient 
$\g_0/2$ is equal to $\omega_I$, so also $\g_0/2$ increases
monotonically with $n$.
Thus, in terms of the equivalent harmonic oscillators,
the least damped mode, which is still the $n=1$ mode, 
is also the one with the 
{\em lowest} value of $\omega_0$, and 
the larger is $(\omega_0)_n$, the shorter is the lifetime, as we
expected from physical intuition.

For large $n$, using \eq{asy} and introducing the Hawking temperature
$T_H=\hbar/(8\pi M)$, \eq{o0oRoI}
can be written in a very suggestive
form,
\be\label{on}
\hbar (\omega_0)_n=\sqrt{m_0^2+p_n^2}\, ,
\ee
where
\be\label{pn}
m_0=T_H \ln 3\, ,\hspace{5mm}
p_n=2\pi T_H  \(n+\frac{1}{2}\)\, .
\ee
The expression for $p_n$ is especially intriguing, since it corresponds to a
particle quantized with antiperiodic boundary conditions on a circle of
length $L=\hbar/T_H=8\pi M$. It is also interesting
to observe that the equal spacing of the
levels for large $n$ is just what would be expected from a description of the
horizon in terms of an effective membrane~\cite{MMmemb}.
We can now 
reexamine some aspects of quantum BH physics, that have been previously
discussed assuming  that the relevant frequencies 
were  $(\omega_R)_n$, using  $(\omega_0)_n$ instead.

{\em Area quantization}.
The idea that in quantum gravity 
the area of the BH horizon is quantized
in  units of $\lpl^2$ (where $\lpl$ is the Planck length) has a long history,
that goes back to Bekenstein~\cite{Bek}. His result was that the area quantum
of a \Sch BH 
is $\D A=8\pi\lpl^2$.  Hod~\cite{Hod} found a
similar quantization, but with a different numerical coefficient, using the
properties of quasinormal modes of \Sch BHs. Since for a \Sch BH 
the horizon area $A$ is related
to the mass $M$ by $A=16\pi M^2$, a change $\D M$ in the BH mass produces a
change $\D A=32\pi M\D M$ in the area. Hod considered a
transition from an unexcited BH to a BH in a mode with $n$ very large. He
argued that for
large $n$ Bohr's correspondence principle should hold, so a semiclassical
description should be adequate even in the absence of a full theory of quantum
gravity, and concluded  from \eq{asy}
that the minimum quantum that can be absorbed in this
transition has $\D M=\hbar\omega =\hbar \ln(3)/(8\pi M)$. This gives 
$\D A=4\ln (3)\lpl^2$ (recall
that in  units $G=c=1$ we have $\lpl^2=\hbar$). 
The numerical factor $4\ln (3)$ generated some excitement because
of possible connections with loop quantum gravity~\cite{Dreyer}
(see however~\cite{DL}).

This conjecture is stimulating, but suffers  from a number of
difficulties. First of all, further studies showed that the factor $4\ln 3$,
that has its origin in $\omega_R$, see \eq{asy}, is not universal. 
For instance,
we can consider a rotating BH with angular momentum per unit mass $a$. 
Computing
the asymptotic behavior of the quasinormal mode frequencies of gravitational
perturbations, 
one finds that  the large $n$ limit and the limit $a\ra 0$ 
do not commute. If we first consider the asymptotic value for a Kerr BH and
then we let  $a\ra 0$, $\omega_R$ does not reduce to
$\ln 3/(8\pi M)$,  but rather vanishes as $a^{1/3}$
\cite{Motl:2003cd,Berti:2003zu,Berti:2003jh,Keshet:2007nv}.
This means that the area quantum becomes arbitrarily
small if we give to the BH an infinitesimal rotation. Similarly, studying
charged BHs, one finds that $\omega_R$ changes discontinuously 
if we interchange
the limits $Q\ra 0$ and $n\ra\infty$. 
Furthermore, the study of generic spin-$j$ perturbations revealed that the
leading 
asymptotic value of the quasinormal mode frequencies is given 
by~\cite{Motl:2003cd}
\be\label{cospij}
e^{8\pi M\omega}=-(1+2\cos \pi j)\, .
\ee
For gravitational perturbations ($j=2$) and for scalar perturbations ($j=0$)
the right-hand side of \eq{cospij} is equal to $-3$, and
we recover \eq{asy}.  For 
vector perturbation ($j=1$), the right-hand side of \eq{cospij} is 
equal to $+1$, and we get
$8\pi M\omega_n = 2\pi i n$,
so the real part is now zero rather than $\ln 3$, and the corresponding quantum
of area would also be zero. Equation~(\ref{cospij}) holds
also for half-integer perturbations \cite{KR,Cho}; in this case 
the right-hand side of \eq{cospij} is 
equal to $-1$, and $8\pi M\omega_n = 2\pi i (n+1/2)$, so again 
${\rm Re}\, \omega_n$ vanishes asymptotically. The conclusion is that 
the asymptotic value of ${\rm Re}\, \omega_n$ (and also whether $p_n$ in \eq{pn}
is proportional to $n$ or to $n+1/2$)
depends on the spin of the
perturbation, and is not an intrinsic property of the BH. 
A similar non-universal behavior was discussed
in \cite{Kuns} in a large class of BH models, that in the $(r_*,t)$ plane
effectively reduce to 2-d dilaton gravity.
In conclusion, the area
quantization determined by Hod's conjecture does not reflect any
intrinsic property of the BH, and the would-be
area quantum vanishes in various
instances.

Another criticism that can be raised to the above derivation
is that one has considered only transitions from the ground state (i.e.
a BH which is not excited) to a state
with large $n$ (or viceversa). However, it is also legitimate 
to consider transitions $n\ra n'$ where
$n$ and $n'$ are both large. Bohr correspondence principle, that
was advocated above, actually only holds for transitions where both 
$n, n'\gg 1$, so
these are in fact the only transitions
that should be considered within the above logic. Now, 
if we use \eq{asy}, we see that
in a transition $n\ra n'$ with $n, n'\gg 1$, 
${\rm Re}\, \omega_n$ changes by 
$O(1/n^{1/2})$. This means that in these transitions the area changes by
arbitrarily small amounts. So, even restricting to 
$j=2$ perturbation of \Sch BHs, the
area quantization
holds only for a transition from (or to) a
BH in its fundamental state, while transitions among excited levels do not
obey it.

All the above difficulties are removed when,  
in Hod's conjecture, we use $(\omega_0)_n$ rather than
$(\omega_R)_n$.
We consider a transition $n\ra n-1$, with $n$ large.
Then $(\omega_0)_n\simeq (\omega_I)_n$ and \eq{asy} gives
an absorbed energy $\D M=\hbar [(\omega_0)_n-(\omega_0)_{n-1}]
=\hbar/(4M)$, so
\be\label{8pi}
\D A= 32\pi M\D M = 8\pi \lpl^2\, ,
\ee
which coincides with the old Bekenstein result. At large $n$
all other transitions require a larger energy; e.g.  $n\ra n-2$
takes away about twice the energy, since for large $n$ the 
$(\omega_0)_n$ are equally spaced. Even if we dare to
extrapolate to low $n$, where semiclassical reasoning might go wrong, we
still remain with a non-vanishing area quantum, of the order of
$8\pi \lpl^2$. As it is clear from Fig.~\ref{fig2}, the transition from
$n=2$ to $n=1$ is the one with the smallest possible jump.
Using the values of $\omega_R$ and $\omega_I$ given in
\cite{Berti}, we find 
$(\omega_0)_{n=2}-(\omega_0)_{n=1}\simeq 0.2/(4M)$, corresponding to
$\D A\simeq 0.2\, (8\pi\lpl^2)$, while
the  transition from $n=1$ to an unexcited BH has
$\D A \simeq 1.5\, (8\pi\lpl^2)$.

Contrary to what happens for $\omega_R$,
the quantum of area obtained from the asymptotics of
$(\omega_0)_n$ is an intrinsic property of \Sch BHs: 
for large $n$ the leading
asymptotic behavior of
$\omega_0$ is given by the $O(n)$ term in $\omega_I$, and 
it does not depend on the spin
content of the perturbation, as we see from \eq{cospij}. Furthermore, 
in contrast to what happens to 
$\omega_R$, for $\omega_I$ the limits $a\ra 0$ and $n\ra \infty$ commute,
and similarly for the limits $Q\ra 0$ and $n\ra \infty$
\cite{Motl:2003cd,Berti:2003zu,Berti:2003jh,Keshet:2007nv}.
The result (\ref{8pi}) can therefore be consistently taken as an indication
of a quantization of the area of the horizon of a \Sch BH. (The generalization
of these results to other spacetimes might however be non-trivial, see 
e.g. \cite{Natario:2004jd}.)
In this context, 
it is useful to
remark that a  gedanken experiment with black holes reveals the existence of a
generalized uncertainty principle, which implies a minimum length of order
$\lpl$ \cite{Maggiore:1993rv}, and which fits very nicely with the above
result.

{\em Entropy and microstates}. If, for large $n$, the horizon
area is quantized, with a quantum $\D A=\a \lpl^2$ (where for us $\a=8\pi$
while for Hod $\a =4\ln 3$), the total horizon area $A$  must be of the form
$A=N\D A=N\a\lpl^2$, where $N$ is an integer. Observe that $N$ is {\em not}
the same as the integer $n$ that we used to label the quasinormal mode. Even
for a BH in its ground state, $n=0$, $N$ is very large since it must account
for the area of the unexcited BH, $N=A/\D A=16\pi M^2/(\a\lpl^2)$. The famous
Bekenstein formula associates to the BH an entropy $S=A/(4\lpl^2)$, and
therefore at level $N$ we expect that a BH should  have a number of
microstates $g(N)$ given by
$g(N)\propto \exp\{(\D A) N/(4\lpl^2)\}=\exp\{\a N/4\}$.
One might try to restrict the possible values of $\a$ as 
follows~\cite{Muk,BekMuk}. One
admits the presence of a subleading term in the Bekenstein formula,
$S=A/(4\lpl^2)+{\rm const.}$, and  fixes the constant requiring that, for
$N=1$, there is only one microstate, $g(N)=1$. This gives
$g(N)=\exp\{ (\a/4) (N-1)\}$. One  then requires 
that $g(N)$ be an integer. This
restricts $\a$ to be of the form $\a = 4\ln k$, with $k$ an
integer. The value $\a = 4\ln 3$ is of this form, which is not the case
for $\a=8\pi$.

However, a number of objections can be raised to this attempt to restrict
$\a$. First of all, in the semiclassical regime where our results could
be trusted, $N$ is very large, of order of $A/\lpl^2$, so
$g(N)$ is the exponential of a a very large number. Even if the number of
microstates must be an integer, there is no hope that a semiclassical
computation can reproduce this number with a precision of order
one, necessary to distinguish an integer from a non-integer value. In fact,
this does not happen even in classical textbook computations in statistical
mechanics. 
Furthermore, the above expression for $g(N)$ 
assumes that the same area quantum $\D A$ is valid from
large $N$ down to $N=1$, where our semiclassical approximation is certainly
unjustified. Indeed, we see from \eqs{on}{pn}  that the levels are equally
spaced only in the limit of highly excited modes, 
otherwise there are deviations. 

Using our value of $\a$ in $S=\a N/4$ we find,
to leading order in the large $N$ limit, 
\be\label{S2piN}
S=2\pi N +O(1)\, ,
\ee
and  $g(N)\propto \exp\{2\pi N\}$. It is quite interesting to observe that 
\eq{S2piN} agrees with the result found in 
Refs.~\cite{Barvinsky:1996bn,Barvinsky:2000gf}, with apparently very different
arguments. In these works, using the periodicity of the {\em Euclidean} BH
solutions, it was found that the entropy is an adiabatic invariant, with a
spectrum given, through Bohr's correspondence principle, precisely by
\eq{S2piN}. This argument required only standard rules 
of quantum mechanics, but
it was somewhat speculative in that the rules were applied in
Euclidean time.

On the other hand,
the periodicity of the Euclidean solution  also entered
implicitly our arguments, since it is at the basis of the analytic
computation of the
asymptotic quasinormal modes frequencies, \eq{cospij}. So it appears
that the periodicity of the BH solutions
in Euclidean time, beside providing a quick derivation of
the value of Hawking temperature, is also  at the origin
of  the area
quantization law.

\vspace{2mm}\noindent
{\em Acknowledgments}. This work  is supported by the Fond National Suisse.
I thank a referee for useful comments.

\end{document}